\documentclass[final]{aipproc}

\layoutstyle{6x9}

\newcommand{\lp}{\left(}
\newcommand{\rp}{\right)}

\newcommand{\ba}{\begin{eqnarray}}
\newcommand{\ea}{\end{eqnarray}}
\newcommand{\be}{\begin{equation}}
\newcommand{\ee}{\end{equation}}

\newcommand{\al}{\alpha}
\newcommand{\bt}{\beta}

\newcommand{\sa}{\sigma}

\usepackage{graphicx}

\begin{document}

\title{Modified Entropic Gravity and Cosmology}
\author{Miguel Zumalac\'arregui}{
address={Institut de Ciencies del Cosmos (ICC-IEEC), Universitat de Barcelona \\ 
Marti i Franques 1, E-08028 Barcelona, Spain (miguelzuma@icc.ub.edu)}
}

\classification{98.80.-k, 04.50.Kd, 04.60.Bc}
\keywords{Cosmology,
Modified theories of gravity
}

\begin{abstract}
It has been recently proposed that the gravitational interaction can be explained as an entropic force. Although a well defined fundamental description for such a mechanism is lacking, it is still possible to address the viability of phenomenological models of entropic-inspired modified gravities. I will summarize some recent work directed to using cosmology as a tool to constraint scenarios aimed to explain the physics behind dark energy and inflation. A phenomenological modification is able to explain cosmic acceleration at the background level and fit observations, but simple inflationary models with higher curvature corrections are incompatible with a matter dominated expansion era.
\end{abstract}

\maketitle

\section{Introduction}

A recent proposal by Verlinde, stating that gravity may be an entropic force \cite{Verlinde:2010hp}, provides a concrete idea to examine the connections between gravitation and thermodynamics \cite{Jacobson:1995ab,Padmanabhan:2009vy} and the emergence of space time \cite{Sindoni:2011ej}. The formulation is heuristic, based on some notions of holography, the assumption of a statistical force $F\Delta x=T\Delta S$, together with some basic thermodynamical relations. Although it has raised several lines of criticism, the idea is difficult to rule out precisely because of the lack of a concrete formulation.

Rather than trying to guess the fundamental underlying mechanism describing the micro-structure of space time, I will describe some approaches to address simple models of entropic inspired modifications of gravity, and use cosmological data to constraint them \cite{Koivisto:2010tb}. In particular, I will focus on models for inflation and dark energy based on  corrections to the horizon entropy-area law of the form
\begin{equation}\label{mastereq}
S=\frac{A}{4\ell_P^2} + s(A)\,.
\end{equation}

\section{Modified Friedmann Equations}

In Verlinde's approach, equation (\ref{mastereq}) can be easily used to derived the modified Newtonian force $F=-\frac{G M m}{R^2}\lp 1 + 4\ell_P^2\frac{\partial s}{\partial A}\rp$. It is then possible to use semi-general-relativistic arguments to construct the modified Friedmann equations that govern the zero order, homogeneous expansion of the universe \cite{Sheykhi:2010yq}.
In the following I will consider a phenomenological power law parameterization of $s(A)$ as well as the leading order corrections expected in some fundamental theories.

\subsection{Phenomenological modifications to the entropy-area law}

\begin{figure}[t]
\includegraphics[width=.45 \textwidth]{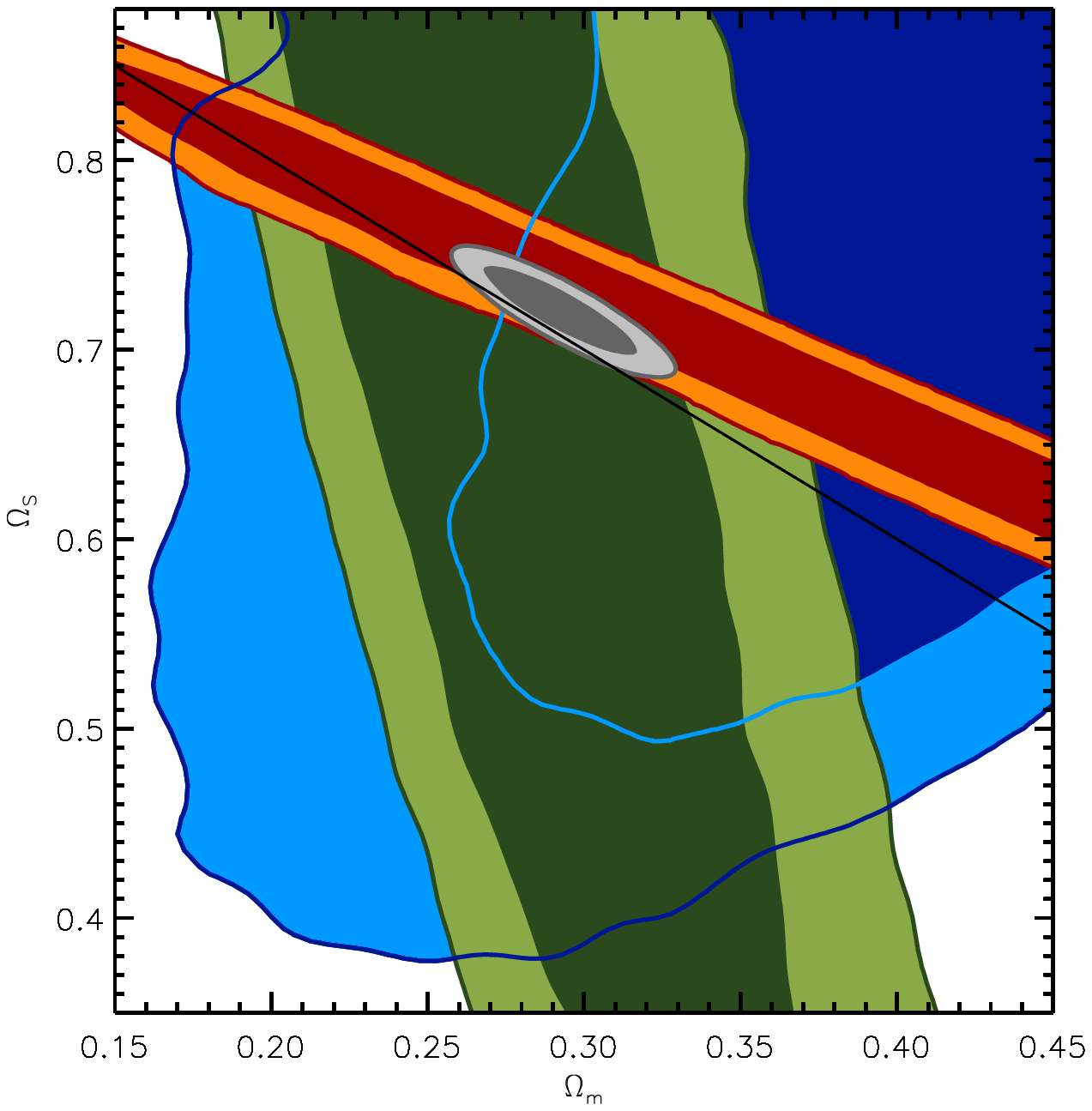} 
\hspace{0.05\textwidth}
\includegraphics[width=.45 \textwidth]{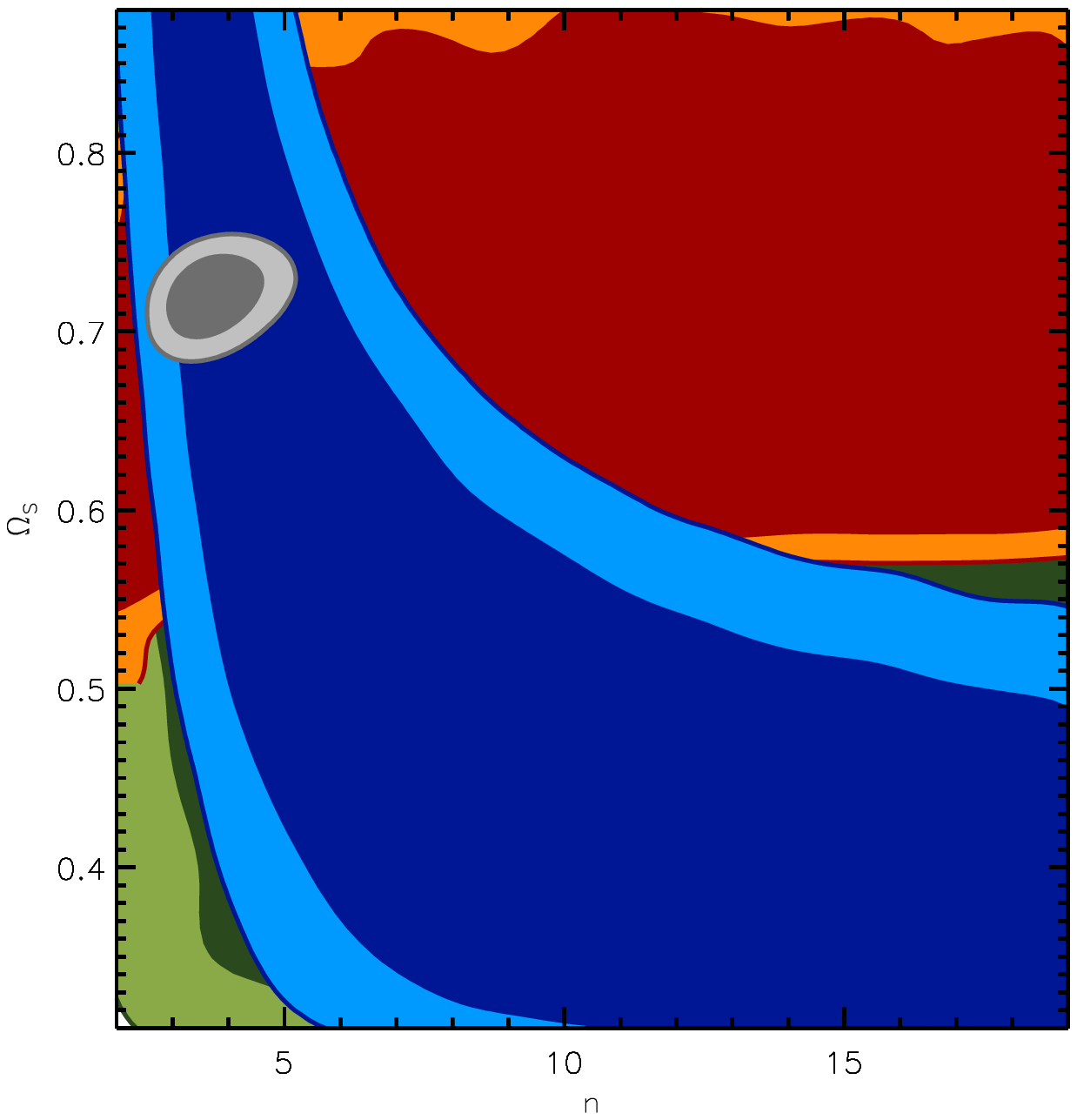}
\caption{Bounds on parameters for power law area-entropy corrections \cite{Koivisto:2010tb} obtained using Sne \cite{union2} (blue), CMB distance priors \cite{wmap} (orange), and BAO \cite{bao7} (green) and the combined datasets (gray). The entropic corrections are parameterized in terms of $\Omega_S$, which is given by the part of equation (\ref{friedmann}) proportional to $\sigma$. Entropic corrections give a slightly worse fit $\chi^2_S = 533.28$ than a similar MCMC for $\Lambda$CDM allowing curvature $\chi^2_\Lambda = 532.34$.}\label{mcmc1}
\end{figure}

A simple way to parameterize deviations from the entropy area law (\ref{mastereq}) is by the addition of a term which is a power law of the area, $s(A) = \frac{4\pi\sa}{n}\lp\frac{A}{\ell_P^2}\rp^{n}$ \cite{Sheykhi:2010yq,Karami:2010bg,Casadio:2010fs}. The modified Friedmann equation in such models has the following form
\begin{equation}
 \label{friedmann}
H^2  +  \frac{k}{a^2} =  \frac{8\pi G}{3}\sum_i \Bigg[ 1  + 
\sa\frac{(1+3w_i)}{1+3w_i-2n}\lp\frac{1}{\ell_P^2(H^2+\frac{k}{a^2})}\rp^{n-1}\Bigg]\rho_i \,.
\end{equation}
Not surprisingly, the low curvature $n>1$ corrections are precisely those which could be significant in cosmology at late times. In the spatially flat case with a single fluid satisfying $p=w\rho$, and when the correction terms dominate in Eq. (\ref{friedmann}), the expansion is described by the effective EoS $w_{eff} = \frac{1+w}{n}-1$, allowing accelerated expansion if $n>3$ (for $w=0$). With larger $n$, the effective EoS is more negative, but phantom expansion can be achieved only when $w$ is itself negative. A Monte Carlo Markov Chain investigation of the parameter space (Figure \ref{mcmc1}) shows that a good fit was obtained for $\Omega_S = 0.69\pm 0.02 $, $n= 3.8 \pm 0.7 $ with a tendence towards a slightly closed universe.

A simple way to study the growth of perturbations is by linking the evolution of an spherical inhomogeneity to the background expansion by assuming that the Jebsen-Birkhoff theorem holds \cite{Lue:2003ky}. In that case, linear perturbations are given as the solutions to the differential equation $\ddot{\delta} + 2H\dot{\delta} = \left(2\dot{H}+\frac{\ddot{H}}{H}\right)\delta$. The r.h.s. side is different to the standard gravity source term $4\pi G \rho_M\delta$, leading to distinct predictions at the onset of late time acceleration. Although current data is unable to significantly sharpen the constraints shown in Figure \ref{mcmc1}, it is foreseeable that future galaxy surveys do.

\subsection{Leading corrections in quantum gravity}

Besides the purely phenomenological parameterization discussed above, some authors have proposed modifications of the entropy-area law based on corrections from quantum gravity, which are argued to be proportional to the logarithm and the inverse of the area \cite{Sheykhi:2010yq}. Another set of proposals link the modifications to the action of the holographic, surface terms in the gravitational action \cite{Easson:2010xf,Easson:2010av}. Neglecting spatial curvature, all these modifications can be encompassed in a set of modified Friedmann equations described by six parameters:
\begin{eqnarray} \label{efs_friedmann}
 H^2 & = & \frac{8\pi G}{3}\rho + \frac{\Lambda}{3} + \al_1H^2+\al_2\dot{H}+ 8\pi G\al_3 H^4\,, \label{efs_1} \\
\dot{H} + H^2 & = & -\frac{4\pi G}{3}(1+3w)\rho +\frac{\Lambda}{3} + \bt_1 H^2 + \bt_2 \dot{H} + 8\pi G\bt_3 H^4\,. 
\label{efs_friedmann2}
\end{eqnarray}
Solutions as a function of the scale factor exist provided that the energy density is described by a fluid with constant equation of state. The cosmological constraints that can be obtained for this type of models are summarized in the following paragraphs, further details can be found in \cite{Koivisto:2010tb}.

\paragraph{Primordial Nucleosynthesys} Neglecting the cosmological constant, the solution of (\ref{efs_friedmann},\ref{efs_friedmann2}) can be recasted in the standard form, but with a modified gravitational constant. Stringent bounds on $\delta G_{eff}/G$ from the abundances of light elementes \cite{Bambi:2005fi} require that $-3.5\cdot 10^{-3} < 2-\frac{(3-\al_1)(1+w)-2\bt_1}{2-\al_2(1+3w)-2\bt_2} < 1.1\cdot 10^{-3} $ 
and $-2 \cdot 10^{84} < \frac{\al_3(1+3w)+2\bt_3}{2-\al_2(1+3w)-2\bt_2} < 6\cdot 10^{84}$. The constraints on the $H^4$ terms ($\alpha_3,\beta_3$) are very weak because of the enormous hierarchy between the Planck and the nucleosynthesis scales.

\paragraph{Inflation} In the limit $a\to 0$, the early time solution yields a constant expansion rate 
$8\pi G H_I^2\approx \frac{(3-\al_1)(1+w)-2\bt_1}{8\pi G(\al_3(1+3w)+2\bt_3)}$. Therefore this model can provide inflation if higher curvature corrections $\alpha_3,\beta_3 \neq 0$ are present (However, this type of corrections enter in conflict with the addition of a cosmological constant, see below). Assuming that inflationary perturbations are generated the same way as by a single scalar field, a tentative value for the $H^4$ corrections can be obtained from the CMB normalization and the spectral index $8\pi G H_I^2 \sim 10^{-10}\epsilon\approx 10^{-12}$. 

\paragraph{Cosmological Constant} For $\Lambda\neq 0$, the solution of (\ref{efs_friedmann},\ref{efs_friedmann2}) has the rather involved form $H^2= \frac{C_1+C_2}{4\pi G} -\frac{2}{1+(a/a_0)^{2C_3}}$, in terms of dimensionless coefficients $C_i(\alpha_j,\beta_j,w,\Lambda)$. This solution fails to provide a matter domination phase and is therefore not cosmologically viable. The particular form is due to the higher order curvature corrections, and since there is no well defined limit $\alpha_3,\beta_3\to 0$, the high order corrections (\ref{efs_friedmann},\ref{efs_friedmann2}).

Once $\alpha_3,\beta_3$ have been set to zero, the effect of the remaining terms is to modify the scaling of matter to an effective e.o.s. $w_M=\frac{2}{3}\left( \frac{(3-\al_1)(1+w)-2\bt_1}{2-\al_2(1+3w)-2\bt_2} -2\right)$. Departures from $w_M=0$ can be constrained by fitting the modified expansion to data from SNe,BAO and CMB distances, resulting in the limit $-17.28 \cdot 10^{-3} < 3 w_M < 20.50 \cdot 10^{-3}$.

Another posibility is to introduce the cosmological constant through a source term, such that the right scaling for presureless matter is preserved. In that situation the dominant effect is a modification of the e.o.s. for $\Lambda$, which is modified to
$ w_\Lambda =\alpha_1 - \beta_1$. The tightest bound on this effect \cite{wmap} is $-0.05 < w_\Lambda < 0.11$.

\section{Conclusions}

The origin of the thermodynamical properties of space-time remains mysterious. Even in that case, simple modified theories of gravity  based on the entropic gravity proposal can be motivated, constructed and explored to a certain extent using cosmological information. Modifications of the entropy-area law can provide mechanisms for dark energy and generically have observable consequences, although simple inflationary scenarios are problematic due to the effects of higher curvature corrections at cosmological constant domination.
It is possible to constraint these simple models even in the absence of a complete formulation, but eventually a full paradigm would be necessary to properly address the full set of predictions and eventually validate or rule out the concept of entropic gravity and its possible variations.\\

{\bf Acknowledgements:} Tomi S. Koivisto and David F. Mota are co-authors of the original work leading to the results presented in these conference proceedings, and their comments were very useful for the elaboration of this manuscript.
I am also thankful to Bruce Bassett and Alvaro de la Cruz-Dombriz for their hospitality at AIMS and UCT while this proceedings were written, as well as Javier Abajo-Arrastia for many enriching discussions.
MZ is funded by MICINN (Spain) through the grant BES-2008-009090.

\bibliographystyle{aipproc}
\bibliography{hrefs}

\end{document}